\documentclass[10pt,letterpaper]{article}
\usepackage{opex3}
\usepackage[]{amssymb,amsmath}
\usepackage{graphicx}
\usepackage{epstopdf}
\usepackage{cite}

\begin{document}
\title{Optimisation of microstructured waveguides in \textbf{\textit{z}}-cut LiNbO$_{\mathbf{3}}$ crystals}
\author{Huseyin Karakuzu,$^1$ Mykhaylo Dubov,$^{1,\ast}$ Sonia Boscolo,$^1$ Leonid A. Melnikov,$^2$ and Yulia A. Mazhirina$^2$ }
\address{$^1$Aston Institute of Photonic Technologies, School of Engineering and Applied
Science,
Aston University, Birmingham B4 7ET, United Kingdom\\
 $^2$State Technical University, 77 Politekhnicheskaya str., 410054,
Saratov, Russia}
\email{$^{\ast}$m.dubov@aston.ac.uk}

\begin{abstract}
We present a practical approach to the numerical optimisation of the
guiding properties of buried microstructured waveguides, which can
be fabricated in a $z$-cut lithium niobate (LiNbO$_3$) crystal by
the method of direct femtosecond laser inscription. We demonstrate
the possibility to extend the spectral range of low-loss operation
of the waveguide  into the mid-infrared region beyond
$3\,\mu\mathrm{m}$.
\end{abstract}

\ocis{ (190.4390) Nonlinear optics, integrated optics; (220.4000)
Microstructure fabrication; (230.7370) Waveguides; (130.5296)
Photonic crystal waveguides; (000.4430) Numerical approximation and
analysis.}

\bibliographystyle{osajnl}


\section{Introduction}

 Tight focusing of intense radiation of femtosecond (fs) laser inside transparent dielectrics
 leads to an alteration of the material properties, e.g., the refractive index (RI), around
 the focal region
 only.
 This enables unprecedented flexibility  in the design of waveguide (WG)
 geometries \cite{Osellame_book2012} and
 the fabrication of various integrated optics components and light-guiding circuits \cite{Suhara_book2003}.
 In crystals, by writing multiple tracks with a slightly reduced RI around the unmodified volume
of material it is possible to produce a depressed-index cladding
\cite{Streltsov_SPIE2003,Okhrimchuk_OL2005} with the central volume
serving as the core of a WG. Tunable, narrow- and broadband light
sources and detectors in the mid-infrared (IR) wavelength range
(2--5$\,\mu\mathrm{m}$) are the focus of intensive research because
of a wide range of enabling applications \cite{Sorokina_book2003},
including ones in remote sensing, spectroscopy, biomedical sensing,
imaging, and others
\cite{Thomas_PSSA2011,Langrockand_JLT2006,Yang_AO2007,Preciado_OL2008}.
 After the advent of quasi-phase matching by poling of
ferroelectric crystals \cite{Armstrong_PR1962}, lithium niobate
(LiNbO$_3$) has become one of the most commonly used materials in
devices such as acousto-optical filters, frequency converters and
optical parametric generators. Microstructured WGs with low losses,
high damage threshold, and controlled dispersion in a broad spectral
range are highly demanded for both traditional applications and
applications in recently emerged fields, such as integrated quantum
optics and mid-IR range frequency comb generation. Therefore, it is
of interest to establish appropriate design principles and develop
suitable fabrication technologies. Recently, depressed-cladding WGs
have been fabricated by fs laser direct writing into bulk optical
glasses (isotropic materials) \cite{Gross_OL2012,Gross_OL2013}. Note
that the numerical model used in these works builds upon a scalar
diffraction approximation and, thus, cannot be used to describe
vector propagation as required by birefringent hosts such as
LiNbO$_3$ crystals.

 In a previous work \cite{Karakuzu_OE2013} we have demonstrated numerically that
 the guiding properties of microstructured WGs in $z$-cut LiNbO$_3$
 crystals
 can be controlled by
 the WG geometry.
 Obviously, with virtually an infinite number of structural parameters the design of such
 structures is not as trivial. Our study ranged over
the parameter space: track size ($D$), period or pitch ($a$), number
of cladding layers ($N_{\mathrm{clad}}$), and RI contrast ($\delta
n$), that is accessible experimentally. In particular, the number of
track rings revealed to play a major role in the control of the
confinement losses, i.e., the losses due to the finite transverse
extent of the confining structure. We observed that for the typical
induced RI contrast $\delta n=-0.01$, increasing the number of
equidistant cladding layers of identical tracks from two to seven in
a microstructured WG with an hexagonal symmetry
 can result in a reduction of the losses in both ordinary (O) and extraordinary (E) polarisations by
 more than three orders of magnitude at the telecommunication wavelengths. Clearly,
 the range
of possible structural parameters that can be investigated  for
 further optimisation is far from being fully explored. We have also
shown \cite{Karakuzu_OE2013} that WG designs with track diameters
that differ from one ring to another
\cite{Viale_COMSOL2005,Renversez_OL2003} can extend the spectral
region of low-loss operation to longer wavelengths. However, it
emerged from our study that a systematic procedure is required to
find
 optimum laws for the variation of WG parameters such as the track size among different rings of tracks.
 Another motivation for the present work is that the RI contrast and the track size are not
 independent parameters -- if the track size varies the RI contrast changes too,
 thus a meaningful optimisation should also account for this dependence. Additionally, the
 direct fs laser inscription \textit{always} generates losses both within and around
 the tracks
 \cite{Turchin_OQE2011,Allsop_AO2010,Aslund_OE2008},
 whose effect should be properly accounted for.
 All the fore-mentioned effects become more significant at the long wavelengths, where we observed
 an unusual variation of
 the confinement losses in the O and E polarisations \cite{Karakuzu_OE2013}.
 In order to confirm the
 accuracy of our numerical model we
 simulated known microstructured optical fibre (MOF) structures
 \cite{Renversez_OL2003}
  using the same material dispersion relations.
 Quantitative agreement for both the dispersion profile and confinement losses
 was obtained within a broad wavelength range.

In this paper, we present a practical approach to the numerical
optimisation of the guiding properties of depressed-cladding, buried
WGs that can be formed in a $z$-cut LiNbO$_3$ crystal by fs laser
writing. The approach accounts for both a suitable variation of the
track size among different track layers, the relationship between
track size and induced RI contrast, and the intrinsic losses due to
fs laser inscription. For our modeling, we used the COMSOL
 simulation software based on the
finite element method (FEM), which enables an accurate description
of both anisotropic materials and materials with
 absorption (indicated by the imaginary part of the RI). To truncate computational grids
 for simulating Maxwell's equations
and, thus, minimize the effect of boundary reflections, we
 used a rather wide perfectly matched layer
absorber surrounding the cross-section of the WG structure
\cite{Karakuzu_OE2013}.
 However, this limits the number of modes that can be followed simultaneously over a wide wavelength
 range. In order to explore the behaviour of a larger number of modes, we
compared our
 method with the plane-wave method (PWM), which has already been used for the analysis of
 anisotropic waveguides \cite{Lu_JOSAA1993} and MOFs
 \cite{Khromova_OC2008,Mazhirina_OS2009,Mazhirina_AIP2010}.
 The PWM can be computationally much faster than the FEM, and it is also well suited for
 the study of mode interactions as well dispersion characteristics changes under external
 influences.
 However, in its current formulation, the PWM cannot provide an accurate
 description of leaky WGs,
  for which the imaginary parts of the effective mode indices are large.

\section{Numerical model for FEM calculations}
\label{sec2}

\begin{figure}[h!]
\centering
\includegraphics[width=0.65\linewidth]{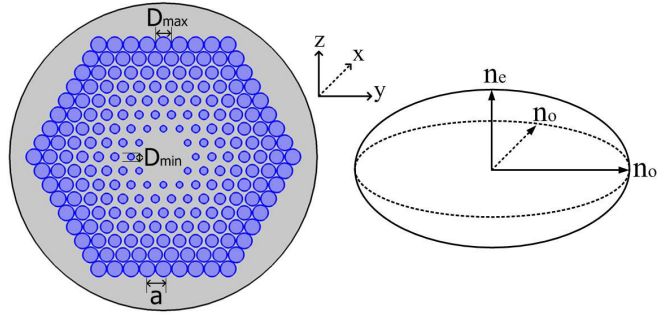}
 \caption{Left: cross-section of WG structure with seven rings of
 tracks with different diameters. Right: ellipsoid of refractive indices of $z$-cut LiNbO$_3$ crystal.
 }
 \label{fig1}
\end{figure}

For this study,  we modeled microstructured WGs with a hexagonal
structure. The depressed cladding was formed by up to seven rings of
cylindrical tracks whose centers were arranged hexagonally, as shown
in Fig.\ 1. These tracks can be written in the bulk of a $z$-cut
LiNbO$_3$ crystal by direct fs laser irradiation using a transverse
inscription geometry.  We found experimentally (the results will be
published elsewhere) that the track diameter $D$
 (in [$\mu\mathrm{m}$]) and the induced RI contrast $\delta n$ depend on the laser pulse energy
 $E$ (in [nJ]) as follows:
\begin{equation}\label{eq1}
\delta n = - 5.1\times 10^{-4}\left( E -
E_{\mathrm{th}}\right),\quad D = 0.193
\left( E -
E_{\mathrm{th}}\right),
\end{equation}
where $E_{\mathrm{th}} = 36.45\,$nJ  is the energy threshold of the
$100\times$ micro-objective used for inscription (numerical aperture
$\mathrm{NA}= 1.25$). These relations were found to be valid for a
laser repetition rate of 11$\,$MHz and the optimum (sample
translation) inscription speed of 10 or 20$\,$mm/s (determined by a
trade-off between inscription depth and laser pulse energy), up to a
maximum available laser pulse energy of approximately $75\,$nJ. It
follows that $\delta n$ can be at most $-0.015$ or $-0.02$. Note
that at laser pulse energies approaching the maximum value the
smoothness of the inscribed tracks may be compromised due to a
strong self-action of the circular laser beam
\cite{Okhrimchuk_LP2009}. In our inscription experiments in
LiNbO$_3$ (which will be described in detail in a future work), RI
contrast reconstruction was performed by quantitative phase
microscopy (QPM) at the optical microscope. The QPM reconstruction
method relies on processing ``$z$-stack'' images of each individual
track by using both commercial and in-house software. This procedure
was calibrated by measuring standard optical fibers with known
specifications. Further, the experimentally observed intensity
distributions of various WG modes were found to be in good
quantitative agreement with the results of numerical modeling.

As mentioned in the introduction, a natural strategy to extend the
spectral range of low-loss operation of the WG structure, is to
allow the track diameter to differ from one ring to another with the
exterior rings that have large tracks. The rate of growth of the
track size from the innermost to the outermost ring can be
parameterized with a single parameter $p>0$, so that the track
diameter $D_n$ in the $n$-th ring is:
\begin{equation}\label{eq2}
  D_n = D_{\mathrm{min}} +\left(\frac{n-1}{N_{\mathrm{clad}}-1}\right)^p  (D_{\mathrm{max}} - D_{\mathrm{min}}),
  \quad n\in[1,N_{\mathrm{clad}}],
\end{equation}
where $D_{\mathrm{max}}$ and $D_{\mathrm{min}}$ are the respective
maximum and minimum diameters. Examples of how the growth rate
parameter $p$ changes the track diameter in a seven-ring WG
structure and the cross-sections of structures for different values
of $p$ are given in Fig.\ 2.

\begin{figure}[htb]
\centering \includegraphics[width=0.525\linewidth]{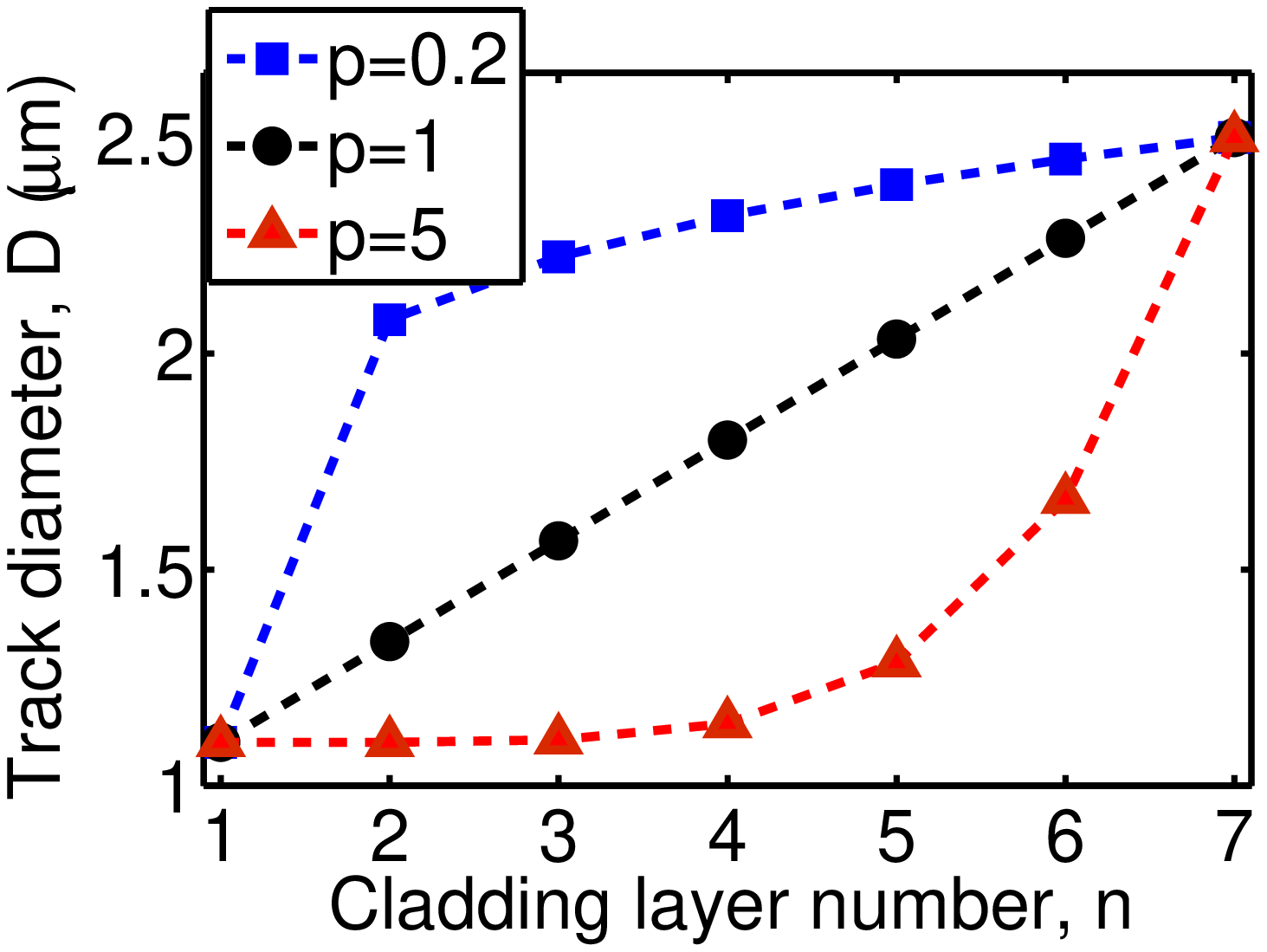}
\includegraphics[width=0.275\linewidth]{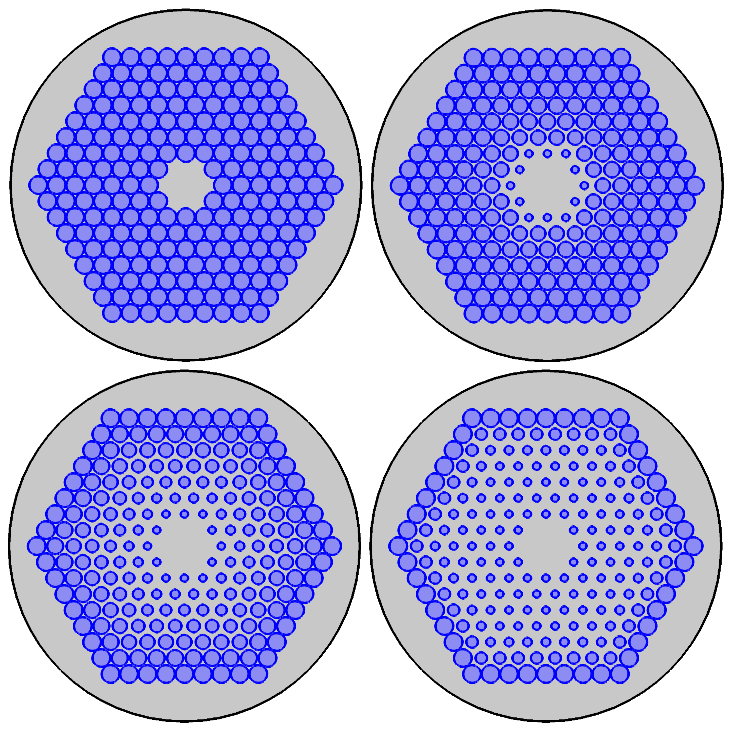}
\caption{Left: track diameter versus cladding layer number for a
seven-ring WG structure with different growth rate parameters $p$.
  Right: Cross-sections of seven-ring structures for (from top to bottom and left to right)
  $p = 0$ (uniform structure), $0.2,\,1,\,5$.
  Other parameters are: pitch $a=2.5\,\mu\mathrm{m}$, $D_{\mathrm{max}}=2.4\,\mu\mathrm{m}$,
  $D_{\mathrm{min}}=1\,\mu\mathrm{m}$.}
  \label{fig2}
\end{figure}

Maxwell's equations  were solved  to find out the complex effective
RIs $n_{o,e}^{\mathrm{eff}}$ of the modes of the structure for the O
and E polarisation states. The confinement losses $\mathcal{L}$ (in
decibels per centimeter) were computed from the imaginary part of
effective RI $\mathcal{I}(n_{\mathrm{eff}})$ through the formula
$\mathcal{L}=40\pi\mathcal{I}(n_{\mathrm{eff}})\times 10^4/[\lambda
\ln(10)]$, where $\lambda$ is given in micrometers. Details of the
numerical computation can be found in \cite{Karakuzu_OE2013}.

\section{Eigenmodes of anisotropic microstructured WGs by PWM}

Maxwell's equations for monochromatic optical waves of angular
frequency $\omega$ propagating along the $x$ axis of the coordinate
system in a transparent, uniaxial crystal with the $y$ optical axis
and of permittivity $\hat{\epsilon}$ yield the following wave
equation for the magnetic field vector $\vec{H}$:
\begin{equation} \label{eq3}
\nabla\times\left({\hat\eta\nabla\times{\vec H}}\right)=K^2\vec
H,\quad \nabla\cdot\vec H=0,
\end{equation}
where $K=\omega/c=2\pi/\lambda$ is the free-space wave number, and
\begin{displaymath}
\hat\eta\equiv\frac{1}{\hat\epsilon}=\begin{pmatrix} \eta_o &0& 0 \\
0 & \eta_e & 0\\
0 & 0 & \eta_o
\end{pmatrix}, \quad \eta_{o,e}=\frac{1}{n_{o,e}^2}.
\end{displaymath}
Here, $n_{o,e}$ are the O and E RIs of the (intact domain of)
crystal, and the distance- and time-dependence of the
electromagnetic field: $\exp(i\beta x-i\omega t)$ is assumed, where
$\beta=K n^{\mathrm{eff}}$ is the propagation constant of the mode
propagating in the structure. For a crystal hosting a
depressed-cladding WG with the track RI contrast $\delta n<0$, we
can use the approximation introduced in \cite{Karakuzu_ASSL2013} and
express the RIs as:
\begin{displaymath}
n_{o,e}(y,z)=n_{o,e}+\delta n f(y,z),\quad f(y,z)=\left\{
\begin{matrix} 1,&(y,z)\in \rm{tracks}\\
 0,&(y,z)\in \rm{otherwise}
 \end{matrix}\right.
 \end{displaymath}
To apply PW decomposition, the WG structure should be put in a box
and tiled periodically along the $y$- and $z$-axes. The box should
not be too large in order not to include a large area of unmodified
material around the structure and, at the same time, should not be
too small in order to avoid overlapping of modes in adjacent
computational cells. Thus, the size of the box has to be a parameter
for the calculations.

The second equation in Eq.\ \eqref{eq3} can be used to exclude
$H_x(y,z)$ from the first equation and, thus, obtain:
\begin{multline}
\beta^2 H_y(y,z)=\frac{\partial^2 H_y}{\partial y^2}+
\frac{\partial^2 H_y}{\partial z^2}+K^2\frac{1}{\eta_o} H_y+
\frac{1}{\eta_o}\frac{\partial \eta_o}{\partial z} \left(
\frac{\partial H_y}{\partial z}-\frac{\partial H_z}{\partial
y}\right),\\
\beta^2 H_z(y,z)=\frac{\eta_o}{\eta_e}\frac{\partial^2 H_z}{\partial
y^2} +\frac{\partial^2 H_z}{\partial z^2}+K^2\frac{1}{\eta_e} H_z
+\left(1-\frac{\eta_e}{\eta_o}\right)\frac{\partial^2 H_y}{\partial
y\partial z} -\frac{1}{\eta_e}\frac{\partial \eta_o}{\partial y}
\left( \frac{\partial H_y}{\partial z} -\frac{\partial H_z}{\partial
y}\right).\label{eq4}
\end{multline}
Equations \eqref{eq4} can be solved using the PWM
\cite{Khromova_OC2008,Mazhirina_OS2009} to find out the (real)
eigenvalues $\beta^2$ and eigenvectors $\{H_y, H_z\}$. For this
purpose, we used the following expansions in terms of PWs:
\begin{gather}
H_y(y,z)=\sum_{m=-\infty}^{\infty}\sum_{n=-\infty}^{\infty}
H^y_{m,n}\exp{\left[iG^y_{m,n}y+iG^z_{m,n}z\right]},\notag\\
H_z(y,z)=\sum_{m=-\infty}^{\infty}\sum_{n=-\infty}^{\infty}
H^z_{m,n}\exp{\left[iG^y_{m,n}y+iG^z_{m,n}z\right]},\notag\\
\frac{1}{\eta_o(y,z)}=\sum_{m=-\infty}^{\infty}\sum_{n=-\infty}^{\infty}
U^o_{m,n}\exp{\left[iG^y_{m,n}y+iG^z_{m,n}z\right]},\notag\\
\frac{1}{\eta_e(y,z)}=\sum_{m=-\infty}^{\infty}\sum_{n=-\infty}^{\infty}
U^e_{m,n}\exp{\left[iG^y_{m,n}y+iG^z_{m,n}z\right]},\notag\\
{\eta_o(y,z)}=\sum_{m=-\infty}^{\infty}\sum_{n=-\infty}^{\infty}
P^o_{m,n}\exp{\left[iG^y_{m,n}y+iG^z_{m,n}z\right]},\notag\\
{\eta_e(y,z)}=\sum_{m=-\infty}^{\infty}\sum_{n=-\infty}^{\infty}
P^e_{m,n}\exp{\left[iG^y_{m,n}y+iG^z_{m,n}z\right]},
 \label{eq5}
\end{gather}
with unknown complex amplitudes $H^{y,z}_{m,n}$, and
$G^{y}_{m,n}=2\pi m/a$, $G^{z}_{m,n}=2\pi n/a$. Substituting these
expansions into Eq.\ \eqref{eq4} and equalising the coefficients of
the same exponential factors one can obtain the following linear
equations for the modal amplitudes $H^y_{m,n}$, $H^z_{m,n}$:
\begin{gather}
\sum^{\infty}_{m^\prime=-\infty}\sum^{\infty}_{n^\prime=-\infty}
\left(L^{yy}_{m^\prime, n^\prime}H^y_{m^\prime,
n^\prime}+L^{yz}_{m^\prime, n^\prime}H^z_{m^\prime,
n^\prime}\right)=\beta^2 H^y_{m,n},\notag\\
\sum^{\infty}_{m^\prime=-\infty}\sum^{\infty}_{n^\prime=-\infty}
\left(L^{zy}_{m^\prime, n^\prime}H^y_{m^\prime,
n^\prime}+L^{zz}_{m^\prime, n^\prime}H^z_{m^\prime,
n^\prime}\right)=\beta^2 H^z_{m,n},\label{eq6}
\end{gather}
where
\begin{gather*}
L^{yy}_{m^\prime,n^\prime}=-\left[(G^y_{m,n})^2+(G^z_{m,n})^2\right]\delta_{mm^\prime}
\delta_{nn^\prime}+K^2 U^o_{m-m^\prime,n-n^\prime}+i
G^z_{m^\prime,n^\prime} V^o_{m-m^\prime,n-n^\prime},\\
L^{yz}_{m^\prime,n^\prime}=i
G^y_{m^\prime,n^\prime}V^o_{m-m^\prime,n-n^\prime},\quad
L^{zz}_{m^\prime,n^\prime}=-(G^z_{m,n})^2\delta_{mm^\prime}\delta_{nn^\prime}
-(G^y_{m^\prime,n^\prime})^2 W_{m-m^\prime,n-n^\prime},\\
-G^y_{m^\prime,n^\prime}G^y_{m^\prime,n^\prime}
T_{m-m^\prime,n-n^\prime}+K^2 U^e_{m-m^\prime,n-n^\prime}+i
G^y_{m^\prime,n^\prime} V^e_{m-m^\prime,n-n^\prime},\quad
L^{zy}_{m^\prime,n^\prime}=i
G^z_{m^\prime,n^\prime}V^e_{m-m^\prime,n-n^\prime},\\
V^{o,e}_{m,n}=
\sum^{\infty}_{m^\prime=-\infty}\sum^{\infty}_{n^\prime=-\infty} i
G^{z,y}_{m^\prime,n^\prime}P^o_{m^\prime,n^\prime}
U^{o,e}_{m-m^\prime,n-n^\prime},\quad W_{m,n}=
\sum^{\infty}_{m^\prime=-\infty}\sum^{\infty}_{n^\prime=-\infty}P^o_{m^\prime,n^\prime}
U^{e}_{m-m^\prime,n-n^\prime},\\
T_{m,n}=
\sum^{\infty}_{m^\prime=-\infty}\sum^{\infty}_{n^\prime=-\infty}
P^o_{m^\prime,n^\prime}
\left(P^{o}_{m-m^\prime,n-n^\prime}-P^{e}_{m-m^\prime,n-n^\prime}\right).
\end{gather*}
From Eq.\ \eqref{eq6}, one can determine $\beta^2$ and
$H^{y,z}_{m,n}$. The transverse components of the electric field
$E_{y,z}$ and the longitudinal component of the Poynting vector
$S_x$ will be then given by
\begin{gather}
E_y=\frac{\eta_e(y,z)}{\beta K}\left(\beta^2 H_z-\frac{\partial^2
H_z}{\partial z^2}-\frac{\partial H_y}{\partial y\partial
z}\right),\quad E_z=-\frac{\eta_o(y,z)}{\beta K} \left(\beta^2
H_y-\frac{\partial^2 H_z}{\partial y \partial z}-\frac{\partial^2
H_y}{ \partial y^2}\right), \notag\\
S_x=\mathcal{R}\left[E_y^* H_z-E_z^* H_y\right].
\end{gather}

The solution of Eq.\ \eqref{eq6} involves the calculation of Fourier
harmonics of the mode profiles. For arbitrary shapes of the
 tracks of
 the WG structure, numerical methods have to be employed to calculate the Fourier coefficients,
 including, when necessary,
 digital image processing of real structure cross-sections \cite{Melnikov_CLEO2003}.
  In the special
case of cylindrical tracks (circular cross-sections), however,
analytical expressions for the Fourier coefficients can be found.
Let $X$ denote any of $\eta_{o,e}$, $1/\eta_{o,e}$, with
$X(y,z)=X_1$ if $(y,z)\in\mathrm{tracks}$ and $X(y,z)=X_2$ if
$(y,z)\in\mathrm{otherwise}$. For cylindrical tracks of radii $r_i$,
one can obtain:
\begin{equation}
X_{m,n}=\left\{ \begin{matrix} \displaystyle\sum_i\frac{\pi
r_i^2}{S}X_1+
\left(1-\sum_i\frac{\pi r_i^2}{S}\right)X_2, & m=n=0 \\
\displaystyle 2 \pi \sum_i
\exp{\left[iG^y_{m,n}y_i+iG^z_{m,n}z_i\right]}
\frac{r_i^2}{S}\frac{J_1(r_i\sqrt{(G^y_{m,n})^2+(G^z_{m,n})^2})}
{r_i\sqrt{(G^y_{m,n})^2+(G^z_{m,n})^2}}(X_1-X_2),& m,n \neq 0
\end{matrix}\right.
\end{equation}
Here $S$ is the area of the surrounding box, $J_1$ is the Bessel
function of the first kind, and $y_i, z_i$ are the coordinates of
the center of the $i$th track.

\subsection{Numerical results by PWM and comparison with FEM}
For practical applications, microstructured WG structures built of
low-RI-contrast tracks should feature low confinement losses and,
thus, need to have a fairly large number of cladding layers
\cite{Karakuzu_OE2013}. For these calculations, we considered
seven-ring structures with the pitch $a=2\mu\mathrm{m}$, the track
radius $r=0.8\mu\mathrm{m}$ and the track RI contrast $\delta
n=-0.01$. The Fourier coefficients and matrices $U_{o,e}$,
$P_{o,e}$, etc.\ were calculated for values of the indices $m,n$ up
to $\pm 35$, and 578 PWs were used. These PWs included mostly modes
propagating along the $x$-axis. The eigenvalue problem was solved
numerically using the ``cg.f'' program from the ``EISPACK'' package
(NetLib). Figure 3 shows the computed modes of the structure. We can
note an overlap between the O and E polarisations at the wavelength
$1.5\,\mu\mathrm{m}$. This overlap will lead to additional losses in
the E mode due to a perturbation induced in the WG.
\begin{figure}[htb]
\centering{\includegraphics[width=0.6\linewidth]{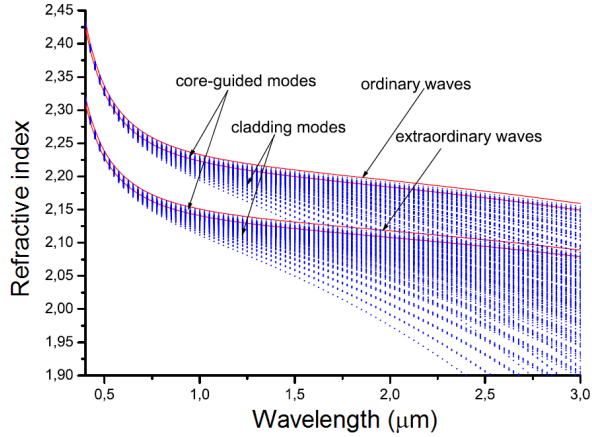}}
\caption{Real parts of effective RIs as a function of wavelength for
the PWM computed modes of a WG structure with seven rings of
tracks.}
\end{figure}
\begin{figure}[htb]
\centering{\includegraphics[width=0.75\linewidth]{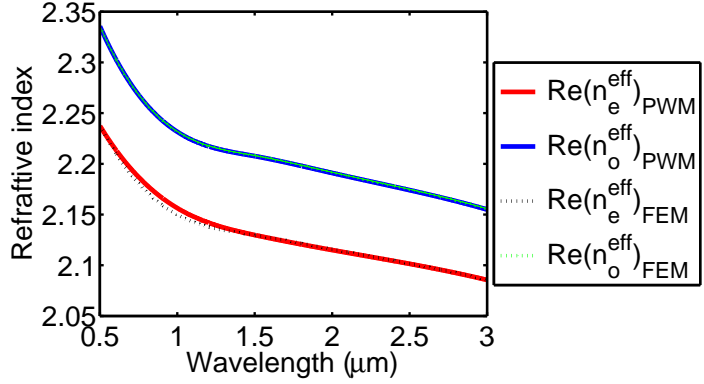}}
  \caption{Real parts of effective RIs for O and E waves as a function of wavelength for the fundamental mode
  of a seven-ring WG structure, as obtained from PWM and FEM simulations.}
  \label{fig4}
\end{figure}

In Fig.\ 4, the effective RI profiles for the fundamental mode are
compared to those obtained using the FEM. The PWM and FEM results
are in excellent agreement for the O polarisation, whereas a small
deviation within the wavelength range $0.5\,\mu\mathrm{m}$ to
$1.5\,\mu\mathrm{m}$ can be observed for the E polarisation. This is
likely due to a leakage of the E mode with lower RI into the
higher-RI O polarisation \cite{Karakuzu_OE2013}, which is not
accounted for by PWM calculations. The effective mode index values
at different wavelengths as obtained from PWM and FEM calculations
are given in Table 1.
\begin{table}[!]
\caption{Real parts of effective mode indices for O and E waves
calculated using the PWM and the FEM. The RIs of the unmodified
material $n_{o,e}$ are also displayed.} \label{tab1}

\begin{tabular}{c|c|c|c|c|c|c}
\hline $\lambda$ ($\mu\mathrm{m}$) & $n_o$ & $n_e$
&$\mathcal{R}\left(n^{\mathrm{eff}}_o\right)_{\mathrm{PWM}}$&
$\mathcal{R}\left(n^{\mathrm{eff}}_e\right)_{\mathrm{PWM}}$
&$\mathcal{R}\left(n^{\mathrm{eff}}_o\right)_{\mathrm{FEM}}$&
$\mathcal{R}\left(n^{\mathrm{eff}}_e\right)_{\mathrm{FEM}}$ \\
\hline 0.5 & 2.33624 & 2.23809 &  2.3354 & 2.2372  &  2.33543   & 2.23728\\
1 & 2.23297 & 2.15087  & 2.23 & 2.1498 & 2.23001& 2.14791\\ 1.5 &
2.20986 & 2.13135 & 2.2039  & 2.129 & 2.20547&2.12545 \\ 2 & 2.19398
& 2.11805 & 2.1847 & 2.1087 &2.18470 &2.10877
\\ 2.5 & 2.17775 & 2.1045   & 2.165   & 2.098   &2.16498 & 2.09173\\ 3 & 2.15921
& 2.08905  & 2.1431  & 2.0758 & 2.14301    & 2.07285\\ \hline
\end{tabular}

\end{table}

\section{Optimisation of WG structural parameters by FEM simulations}

To minimise the confinement losses in the WG structure at the long
wavelengths and, thus, extend the spectral range where the loss
figures for the modes are acceptably low, we performed FEM
simulations. This is because the PWM in its current formulation does
not allow to calculate the imaginary parts of the effective mode
indices. To find the leakage losses with the PWM, one should compute
the localization coefficients for the modes or, alternatively,
include an absorbing medium at the boundary of the computational
supercell. The first approach would require the ``calibration'' of
the losses for given localization coefficients, while the second
approach is similar to using a PML absorber as in the FEM. We assume
here 1$\,$dB/cm to be an acceptable loss level for technological
applications. A simple idea for addressing the numerical
optimisation problem described in this section came from the
observation that the confinement losses become monotonic functions
of wavelength at sufficiently long wavelengths. Thus, one can vary
different WG structural parameters at a fixed wavelength of
interest, and only when the best loss figures are obtained, perform
the full wavelength scan. This makes the optimisation procedure much
less time consuming and, thus, practically feasible. The FEM results
presented here refer to the fundamental mode of the structure, which
was selected using the criterium of 'minimum effective mode area'
during the wavelength scanning \cite{Karakuzu_OE2013}. The RI
contrast induced by fs inscription is the most important parameter
for mid-IR operation (and successful optimisation) of the WG. Figure
5 illustrates the dependence of the confinement loss on the track
diameter at the wavelength $\lambda=1.55\,\mu\mathrm{m}$ for a
seven-ring, uniform ($p=0$) structure with different values of
$\delta n$. It is seen that for $\delta n=-0.01$ and $\delta
n=-0.02$ there is a ``plateau'' of low losses over the diameter
range 2.2$\,\mu\mathrm{m}$ to 2.5$\,\mu\mathrm{m}$ and
1.4$\,\mu\mathrm{m}$ to approximately 2.2$\,\mu\mathrm{m}$,
respectively. Remarkably, such a plateau does not appear for lower
RI contrasts, which are typically obtained by low-repetition rate fs
laser inscription \cite{Okhrimchuk_LP2009}. As we mentioned in Sec.\
\ref{sec2}, the RI contrasts of smooth tracks that can be achieved
in crystals with current fs micro-fabrication technology are around
$-0.01$ or $-0.02$. Higher RI contrasts can still be obtained by fs
inscription, with the fs laser creating severe damage tracks inside
the bulk at the focal volume
\cite{Apostolopoulos_APL2004,Rodenas_OE2011}. Damage tracks,
however, are not smooth and, thus, typically not suitable for
low-loss light guiding.
\begin{figure}[htb]
\centering{\includegraphics[width=0.5\linewidth]{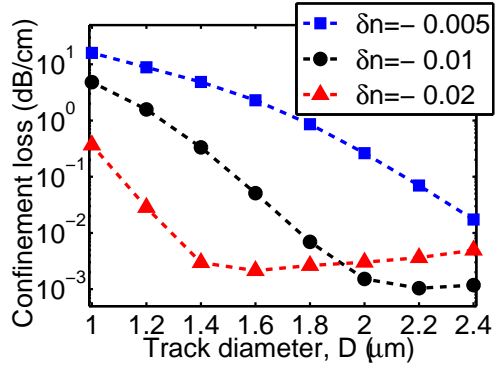}}
\caption{Confinement loss for E wave at
$\lambda=1.55\,\mu\mathrm{m}$ versus track diameter for a seven-ring
WG structure with $a=2.5\,\mu\mathrm{m}$ and different values of
$\delta n$.} \label{fig5}
\end{figure}

The influence of a varying track diameter among the different
cladding layers on the loss properties of the WG at the wavelengths
$1.55\,\mu\mathrm{m}$ and $3\,\mu\mathrm{m}$ is illustrated by Fig.\
6. Our simulations showed that the most interesting range of growth
rate parameter values is $0<p\leq 1$. We can see from Fig.\ 6 that
as $p$ decreases from 1, the confinement loss also decreases up to
much lower values than that of a uniform structure (corresponding to
$p=0$ when $D_n=D_{\mathrm{max}}$ $\forall n$). At
$\lambda=1.55\,\mu\mathrm{m}$, the loss profile is flat over the $p$
range $p\to 0$ (yielding $D_1=D_{\mathrm{min}}$ and $D_n\approx
D_{\mathrm{max}}$, $n>1$) to $p=0.5$ for the maximum track diameter
$D_{\mathrm{max}}=2.4\,\mu\mathrm{m}$. For a uniform structure
$D_{\mathrm{max}}=2.2\,\mu\mathrm{m}$ yields lower confinement loss
than $D_{\mathrm{max}}=2.4\,\mu\mathrm{m}$. On the other hand, the
loss for a structure with $p=1$ is higher than that for the uniform
structure at $D_{\mathrm{max}}=2.2\,\mu\mathrm{m}$. Differently, at
$\lambda=3\,\mu\mathrm{m}$ the decrease of the confinement loss with
decreasing $p$ values is approximately linear for both
$D_{\mathrm{max}}=2.2\,\mu\mathrm{m}$ and
$D_{max}=2.4\,\mu\mathrm{m}$. The smallest loss value
$\mathcal{L}=0.5\,$dB/cm is obtained for $p$ very close to 0 and
$D_{\mathrm{max}}=2.4\,\mu\mathrm{m}$. The possibility of achieving
such low loss figures at $\lambda=3\,\mu\mathrm{m}$ makes the WG
suitable for mid-IR applications. Note that the transparency region
of the WG can be further extended to longer wavelengths by
increasing $D_{\mathrm{max}}$ and properly fitting the $p$ parameter
to the RI contrasts being used.
\begin{figure}[htb]
\centering{\includegraphics[width=0.45\linewidth]{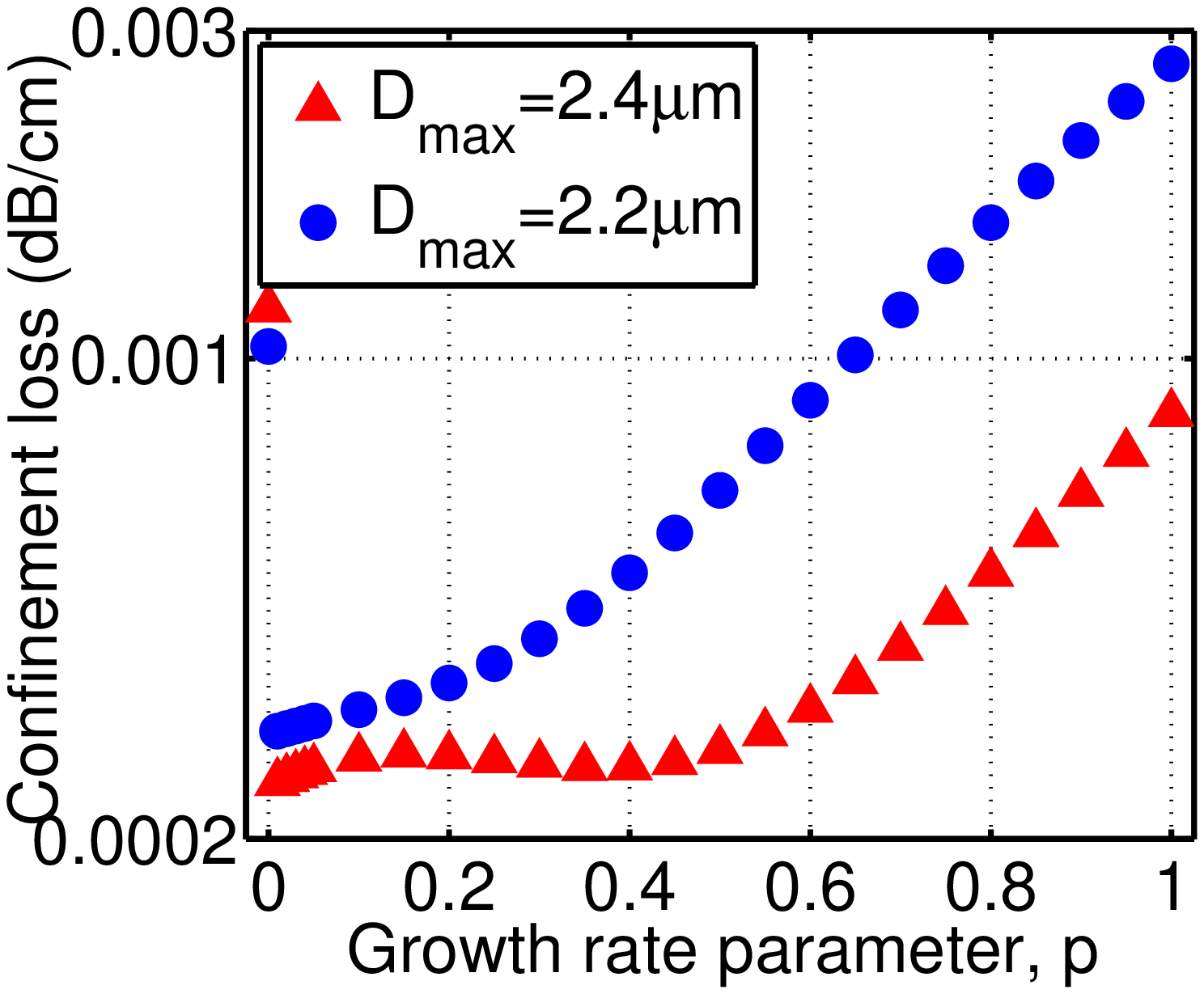}
\includegraphics[width=0.45\linewidth]{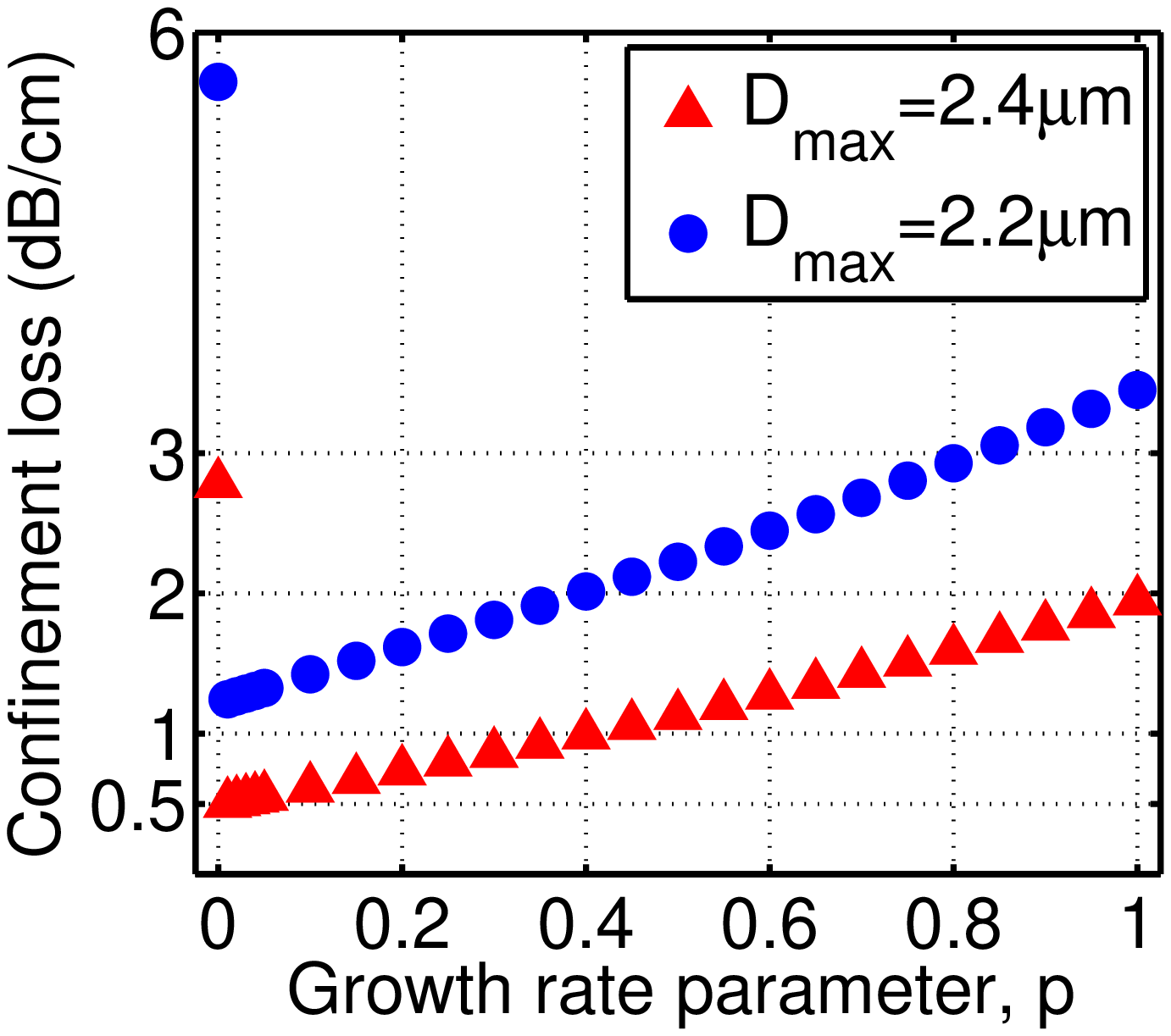}}
\caption{Confinement loss for E wave versus growth rate parameter
$p$ at $\lambda=1.55\,\mu\mathrm{m}$ (left) and
$\lambda=3\,\mu\mathrm{m}$ (right), for a seven-ring WG structure
with $\delta n = -0.01$. Here, $D_{\mathrm{min}}=1\,\mu\mathrm{m}$.
}\label{fig6}
\end{figure}

Figure 7 shows the variation of the confinement loss across the
wavelength range 0.3$\,\mu\mathrm{m}$ to 3$\,\mu\mathrm{m}$ for WG
structures with various growth rate parameters. One can see that
while for a uniform structure the losses in both O and E
polarisations are below 1$\,$dB/cm for wavelengths up to
approximately  $1.8 \,\mu\mathrm{m}$, the spectral range where the
losses are acceptably low is extended up to
$\lambda=3\,\mu\mathrm{m}$ for a structure with $p=0.01$. The
optimisation of the growth rate parameter results in a reduction of
the losses in both polarisations by two orders of magnitude at
$\lambda=3 \,\mu\mathrm{m}$. One may also notice that while the
confinement losses of a WG with $p= 0.5$  are lower than those of a
WG with $p=1$ for wavelengths below $2.6 \,\mu\mathrm{m}$, a
reversal of trend happens at the wavelengths above
$2.6\,\mu\mathrm{m}$. Further, Fig.\ 6 highlights the distinctly
different behaviours of the confinement losses in the O and E
polarisations at the low wavelengths. The confinement loss for the E
wave stops decreasing with decreasing wavelength below some
wavelength which is specific to the WG structure, and features
resonance effects. The critical wavelengths below which such
behaviour is observed are $1.8\,\mu\mathrm{m}$,
$1.6\,\mu\mathrm{m}$, $1.4\,\mu\mathrm{m}$ and $1.2 \,\mu\mathrm{m}$
for WGs with the respective growth rates $p=0.01$, $p=0.5$, $p=1$
and $p=0$. This resonance behaviour which is peculiar to E-polarised
propagating waves in anisotropic WGs, stems from the coupling of the
E-polarized fundamental  mode to the radial modes of O
 polarisation and consequent leakage of the E wave through these modes \cite{Lu_JOSAA1993}.
\begin{figure}[htb]
\centering{\includegraphics[width=0.9\linewidth]{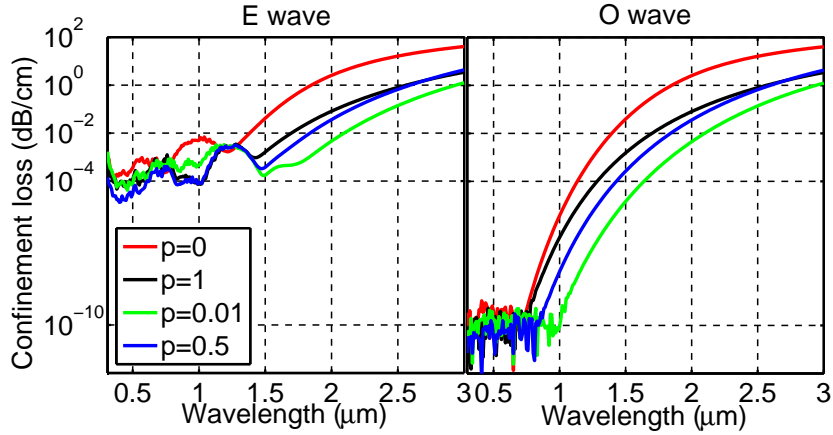}}
\caption{Confinement losses for O and E waves as a function of
wavelength for a seven-ring WG structure with different growth rate
parameters $p$
. Other parameters are: $\delta n = -0.01$,
$D_{\mathrm{max}}\,=$2.2$\,\mu\mathrm{m}$,
$D_{\mathrm{min}}\,$=1$\,\mu\mathrm{m}$. } \label{fig7}
\end{figure}

\begin{figure}[htb]
\centering{\includegraphics[width=0.55\linewidth]{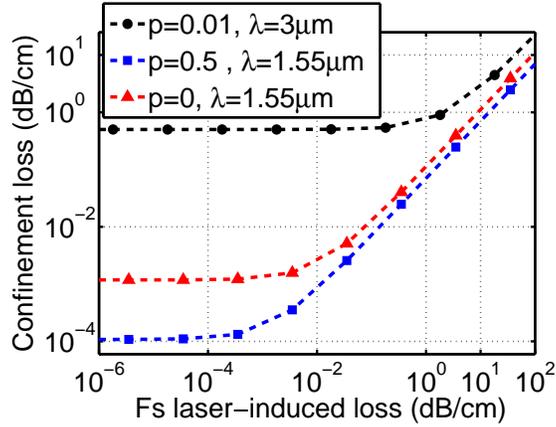}}
\caption{ Confinement loss for E wave versus loss induced on tracks
by fs inscription for a WG with $p=0$ and $p=0.5$ at $\lambda=1.55
\,\mu\mathrm{m}$, and with $p=0.01$ at $\lambda=3 \,\mu\mathrm{m}$.
Other WG parameters are: $\delta n = -0.01$, $a=2.5
\,\mu\mathrm{m}$, $D_{\mathrm{max}}=2.4 \,\mu\mathrm{m}$,
$D_{\mathrm{min}}=1 \,\mu\mathrm{m}$.} \label{fig8}
\end{figure}

The results presented so far were obtained by assuming that the RI
change induced in the material by direct fs laser inscription is a
real value. In fact, as mentioned in the introduction, the fs
irradiation always induces material absorption, which needs to be
accounted for in the WG design. To this end, we computed the
confinement loss in a WG with $p=0$ and $p=0.5$ at
$\lambda=1.55\,\mu\mathrm{m}$, and with $p=\,0.01$ at
$\lambda=3\,\mu\mathrm{m}$ for a range of fs laser-induced loss
values. The results are shown in Fig.\ 8, which reveals that the
higher is the confinement loss of the WG, the lower is the WG
sensitivity to the imaginary part of the induced RI contrast.
Indeed, induced losses of up to 1$\,$dB/cm do not affect the
confinement loss at $\lambda=3\,\mu\mathrm{m}$, whereas the effect
of induced loss is more important at $\lambda=1.55\,\mu\mathrm{m}$,
where the WG exhibits lower confinement loss. We note that the
imaginary part of the modified RI could be measured by using, for
example, the Born scattering interferometry method recently proposed
in \cite{Turchin_OQE2011}.

Finally, we included in our design procedure also the relationship
between induced RI contrast $\delta n$ and track size $D$. Indeed,
as we mentioned before, the dependence of $\delta n$ and $D$ on the
laser pulse energy makes such parameters correlated if the sample
translation speed is fixed (Eq.\ \eqref{eq1}). Note that it is
possible to experimentally trim these parameters to the desired
values by tuning both the laser pulse energy and the sample
translation speed, as both produce albeit connected but not
identical changes to $\delta n$ and $D$. Simulation results are
presented in Fig.\ 9, which shows the variation of the confinement
losses in the O and E polarisations as a function of wavelength for
a seven-ring WG structure with a growth rate parameter of $p=0.01$
and where the RI contrast was changed from one ring of tracks to
another following the change in the track size. It is seen that the
spectral region where the losses in both polarisations are below
1$\,$dB/cm extends up to $3.5\,\mu\mathrm{m}$ for this optimised
structure.
\begin{figure}[htb]
\centering{\includegraphics[width=0.55\linewidth]{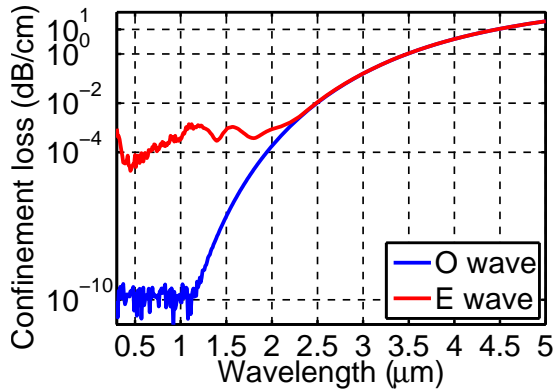}}
\caption{Confinement losses for O and E waves as a function of
wavelength for a seven-ring WG structure with $p=0.01$. Other
parameters are: maximum RI contrast $\delta n=-0.01$,
$D_{\mathrm{max}}=2.4\,\mu\mathrm{m}$,
$D_{\mathrm{min}}=1\,\mu\mathrm{m}$.}\label{fig9}
\end{figure}

\section{Conclusions}
We have presented a numerical approach to the optimisation of the
guiding properties of depressed-index cladding WGs that can be
fabricated in a $z$-cut LiNbO$_3$ crystal by direct fs laser
inscription. The approach accounted for both a suitable variation of
the track size among different cladding layers, the relationship
between track size and induced RI contrast, and the losses induced
on the tracks by fs irradiation. We have shown that the spectral
region where the confinement losses in both O and E polarisations
are acceptably low (below 1$\,$dB/cm) can extend up to a wavelength
of $3.5\,\mu\mathrm{m}$ for optimised, hexagonal WG structures with
seven rings of tracks. This makes such structures suitable for
mid-IR applications. We note that the possibility of further
extending the low-loss operation region of these WGs depends on the
ability to experimentally achieve higher RI contrasts than those
that are feasible by use of current fs micro-fabrication technology,
as well as to increase the track sizes.

\section*{Acknowledgments}
The authors would like to acknowledge support by the Leverhulme
Trust (grant RPG-278) and the Engineering and Physical Sciences
Research Council (grant EP/J010413/1).

\end{document}